\documentclass[twocolumn,pra,showpacs,aps]{revtex4}

\usepackage{amsmath}
\usepackage{amsbsy}
\usepackage{amsthm}
\usepackage{amssymb}
\usepackage{latexsym}
\usepackage[dvips]{color,graphicx}

\begin{document}

\title{Confinement control by optical lattices}

\author{Marcos Rigol}
\affiliation{Institut f\"ur Theoretische Physik III, 
Universit\"at Stuttgart, Pfaffenwaldring 57, D-70550 Stuttgart, Germany.}
\author{Alejandro Muramatsu}
\affiliation{Institut f\"ur Theoretische Physik III, 
Universit\"at Stuttgart, Pfaffenwaldring 57, D-70550 Stuttgart, Germany.}

\begin{abstract}
It is shown that the interplay of a confining potential with a periodic 
potential leads for free particles to states spatially confined on a 
fraction of the total extension of the system. A more complex ``slicing'' 
of the system can be achieved by increasing the period of the lattice 
potential. These results are especially relevant for fermionic systems, 
where interaction effects are in general strongly reduced for a single 
species at low temperatures.
\end{abstract}
\maketitle

\section{Introduction}

The study of trapped atomic gases has become a field of intense research in 
the past years. The realization of Bose-Einstein condensation (BEC) in trapped 
dilute atomic vapors \cite{anderson,bradley,davis} was the main motivation 
starting all the experimental and theoretical research in this area. BEC was 
obtained trapping and evaporatively cooling bosonic alkali metals. Recently, 
the possibility of trapping and cooling Fermi gases has attracted a lot of 
attention, due to the fact that in the quantum degeneracy regime, superfluidity 
appears within reach \cite{ohara}. However, cooling single component Fermi 
gases up to very low temperatures is more difficult than cooling bosonic gases 
since the $s$-wave collisions are forbidden for identical fermions. On the 
other hand, single species Fermi gases make in this way possible to access 
experimentally an ideal Fermi gas. As shown below, such a simple system can 
develop a rich behavior by the combination of a confining and a lattice 
potential.    

For atoms confined in a harmonic trap, a case that adequately describes 
most of the experiments realized so far \cite{dalfovo}, a fairly complete 
theoretical understanding was achieved for the one-dimensional (1D)
\cite{gleisberg,vignolo1,minguzzi,vignolo2} and two- or three-dimensional 
\cite{schneider,brack,vignolo3} single component spin polarized trapped Fermi 
gas, which at very low temperatures can be considered as a noninteracting gas. 
The harmonic form of the potential allows obtaining a number of exact 
analytical results for these systems. However, these results cannot be 
extended to incorporate an additional lattice potential, a case of increasing 
interest after the experimental realization of a Mott insulator in the 
presence of an optical lattice \cite{greiner}. A further interest on the 
introduction of an optical lattice in fermionic systems arises from the 
possible connections with central problems in condensed matter physics 
\cite{zoller}.

We analyze here ground state properties of single species noninteracting 
fermions confined on 1D optical lattices. These systems are relevant for the 
understanding of recent experimental results \cite{modugno03,ott04,pezze04}, 
where due to the very low temperatures achieved, fermions can be considered 
as noninteracting particles. On the theoretical side, the Hamiltonian can be 
diagonalized numerically, which allows to consider any kind of trapping 
potential and any number of dimensions for the system. We show that 
the interplay between the lattice and the confining potential leads in a region 
of the spectrum to a splitting of the system with eigenstates that have a 
nonvanishing weight only in a fraction of the trap. Hence, such systems are 
qualitatively different from the cases without the lattice, which have been 
studied recently 
\cite{gleisberg,vignolo1,minguzzi,vignolo2,schneider,brack,vignolo3}.
 
We also study the nonequilibrium dynamics of the fermionic cloud 
on a lattice. In particular, we study the case in which the center of the 
trap is initially displaced a small distance. It allows to realize the 
existence of the single particle states confined in a part of the trap 
obtained in the equilibrium case, since for some values of the parameters 
the center-of-mass (c.m.) of the system oscillates in one side of the trap. 
With these results, we reproduce the experimental observations in Refs.\ 
\cite{modugno03,ott04,pezze04}, and complement other theoretical approaches 
to this problem \cite{pezze04,kennedy04,ruuska04}.

We show that if in addition to the lattice an alternating potential is 
introduced, doubling the original periodicity, an additional ``slicing'' of the 
system can be achieved. The width and number of such regions can be controlled 
in a given energy range by the amplitude of the new modulation. By filling 
these systems with fermions, insulating regions may appear, that in the case of 
an alternating potential, are similar to the Mott insulating plateaus of the 
trapped fermionic Hubbard model \cite{rigol03_1,rigol03_2}. In the 
noninteracting case it is possible to calculate the local density of states, 
which exhibits the presence of a local gaps in the system. In addition, a 
local compressibility \cite{rigol03_1,rigol03_2} also serves as a local order 
parameter to characterize the insulating regions. This extends the results 
initially obtained for the bosonic case \cite{batrouni}, showing that in 
general, the distinction between commensurate and incommensurate fillings 
typical in extended solid-state systems is lost in the trapped system.

The presentation is organized as follows: In Sec.\ II we study 1D lattices 
superposed to a confining potential. We analyze the generic features valid for 
any kind of trapping potential, and focus on fermionic systems. In Sec.\ III, 
an analysis of the nonequilibrium dynamics of the 1D trapped fermions is 
presented, and recent experimental results reproduced. In Sec.\ IV we study 
the case in which an additional alternating potential in introduced, 
and dicuss analogies and differences with the results obtained 
for the fermionic Hubbard model. In Sec.\ IV, we extend the analysis of 
Sec.\ II and IV to two dimensions (2D). Finally the conclusions are given in 
Sec.\ V.

\section{Noninteracting particles confined in 1D optical lattices}

We analyze in this section 1D noninteracting systems confined by arbitrary 
potentials when an underlying optical lattice is present. We first show results 
for a harmonic confining potential and then discuss the features that are 
generally valid for any other kind of confining potential. For definiteness we 
concentrate on the fermionic case, although the spectral features are equally 
valid for bosons, since we deal with the noninteracting case. 

In the second quantization language, the Hamiltonian describing a confined 
dilute and ultracold (noninteracting) gas of single-species fermions, 
under the influence of a 1D optical lattice, can be written as
\begin{equation}
H=\int d{\bf r}\ \hat{\Psi}^{\dagger}({\bf r})
\left[ -\dfrac{\hbar^2}{2m}\nabla^2+V({\bf r})+V_0(x)
\right] \hat{\Psi}({\bf r}),
\label{HamFOL1}
\end{equation}
where $\hat{\Psi}^{\dagger}({\bf r})$ and $\hat{\Psi}({\bf r})$ are the 
creation and annihilation fermionic field operators, respectively. 
The confining potential is denoted as $V({\bf r})=V(x)+V(y)+V(z)$. We 
analyze in this section the case in which the transversal component of the 
confining potential $V(y)+V(z)$ is very strong so that only its lowest energy 
state is populated, and the exited states are not accessible for the given 
experimental setup. Hence, the relevant dynamics of the system is restricted 
to occur in the longitudinal direction where the trap is considered to have an 
arbitrary power $\alpha$, $V(x)=V_\alpha x^\alpha$. In Eq.\ (\ref{HamFOL1}),
$V_0 (x)=V_{0}\cos^2(k x)$ describes the potential generated by a 1D 
optical lattice. The wave vector $k=2\pi/\lambda$ is determined by the 
wavelength $\lambda$ of the laser beam. (The lattice spacing is then 
$a=\lambda/2$.) Assuming the atoms to be at the 
lowest vibrational level in each site, the fermionic field operators can 
be expanded in single band Wannier functions $\phi_i(x)$,  
$\hat{\Psi}(x)=\sum_i c_{i} \phi_i(x)$, 
and from Eq.\ (\ref{HamFOL1}) one obtains the single band Hamiltonian
\begin{equation}
\label{Hamiltonian} H = -t \sum_{i} \left( c^\dagger_{i}
c^{}_{i+1} + \text{H.c.} \right) + V_\alpha \sum_{i} x_i^\alpha \ n_{i },
\end{equation}
where $c^\dagger_{i}$ and $c^{}_{i}$ are creation and annihilation operators, 
respectively, for a spin polarized fermion on site $i$, the local density is 
$n_{i } = c^\dagger_{i} c^{}_{i }$, and $x_i$ measures the positions of the 
sites in the trap ($x_i=i a$ with $-N/2+1\leq i\leq N/2$, $N$ being the 
number of lattice sites). The hopping parameter is denoted by $t$, which for 
$V_0\gg E_r$ can be written in terms of the experimental parameters as 
$t=4/\sqrt{\pi}(V_0/E_r)^{3/4}E_re^{-2\sqrt{V_0/E_r}}$ \cite{zwerger}, 
where the recoil energy of the atoms (with mass $m$) is $E_r=\hbar^2k^2/2m$. 
The total number of spin polarized fermions in the system is denoted by $N_f$.
We diagonalize the Hamiltonian numerically, and consider the cases in 
which all particles are confined.

\begin{figure}[h]
\includegraphics[width=0.425\textwidth,height=0.44\textwidth]
{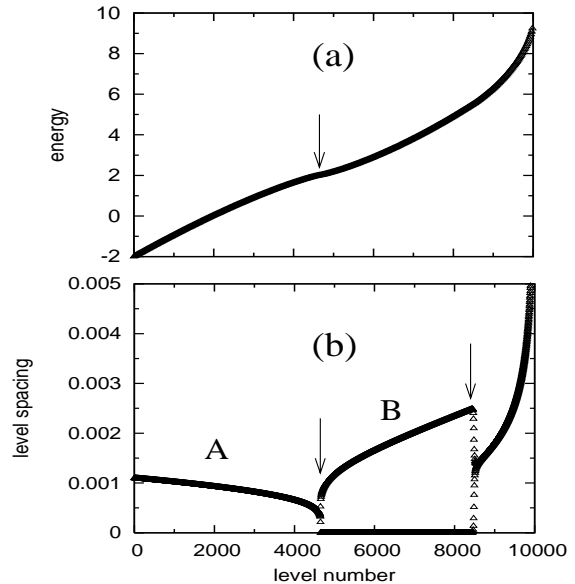}
\caption{Single particle spectrum (a) and level spacing (b) 
for a system with $N=10000$ and $V_2a^2=3 \times 10^{-7} t$. 
Energies are measured in units of $t$.
For the explanation of the arrows see text.}
\label{ELV_I0}
\end{figure}
Results obtained for the single particle spectrum of a system 
confined by a harmonic potential are presented in Fig.\ \ref{ELV_I0}(a). 
The spectrum is clearly different from the usual straight line in the absence 
of a lattice. It is possible to see that in Fig.\ \ref{ELV_I0}(a) the 
spectrum can be divided in two regions according to the behavior of the
energy as a function of the level number. An arrow is introduced
where a change in the curvature is observed. More detailed information 
can be obtained by considering the level spacing 
as a function of the level number [Fig.\ \ref{ELV_I0}(b)]. 
There it can be seen that in the low energy part of the spectrum (region A),
the level spacing decreases slowly with increasing level number, 
in contrast to the case without the lattice in which
the level spacing is constant. However, at the point signaled with
the first arrow, a qualitative change in the single particle
spectrum occurs, characterized by an oscillating behavior of the 
level spacing. The part with values of the level spacing increasing 
with the level number corresponds to odd level numbers and the one with
a level spacing that decreases up to zero corresponds to even
level numbers. That is, a degeneracy sets in that continues up to 
the point signaled with the second arrow, where 
a new change in the behavior of the level spacing shows up.
The region beyond the second arrow corresponds to deconfined states, 
which are of no interest since experimentally they are associated to
particles that scape from the trap (which in the system of 
Fig.\ \ref{ELV_I0} has 10000 lattice sites).

\begin{figure}[h]
\includegraphics[width=0.49\textwidth,height=0.24\textwidth]
{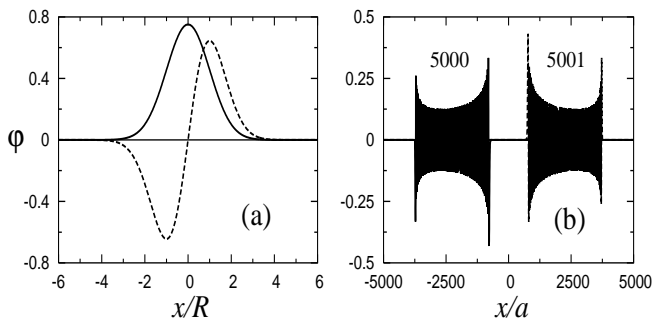}
\caption{Scaled HO orbitals in the presence of a
lattice for $N=10000$ and $V_2a^2=3 \times 10^{-7}t$. 
(a) First (continuous line) and second (dashed line) HO orbitals,
(b) HO orbitals 5000 (only different from zero for negative $x$) 
and 5001 (only different from zero for positive $x$).
In (a) the positions are given in units of the HO length 
$R=\left(V_2/ta^2\right)^{-1/4}$ (for an explanation, see text), and 
in (b) in units of the lattice constant $a$.}
\label{Orbitals}
\end{figure}

In the lowest part of the spectrum of Hamiltonian (\ref{Hamiltonian}),
the eigenfunctions are essentially the harmonic oscillator (HO) orbitals 
in the absence of a lattice. This is shown in Fig.\ \ref{Orbitals}(a) 
for the first and the second eigenfunctions of Eq.\ (\ref{Hamiltonian}), 
and the same parameters of Fig.\ \ref{ELV_I0}.
These orbitals are perfectly scalable independently of the size of the 
system and of the ratio between $V_2$ and $t$. It is only needed to 
consider that the usual HO characteristic length 
$R\sim \left(m\omega \right)^{-1/2}$ (without the lattice) is given in terms 
of the lattice parameters through $R \sim \left(V_2/ta^2\right)^{-1/4}$, 
with the effective mass $m\sim \left(ta^2\right)^{-1}$ for very low energies, 
so that the scaled orbitals are given by 
$\varphi=\left( R/a\right) ^{1/2}\phi$ where $\phi$ are 
the HO orbitals with the lattice, i.e., the same relation 
as for the HO {\em without} the lattice holds 
for the lowest energy orbitals {\em with} the periodic potential.
This implies that very dilute systems behave similarly to continuous 
systems, which have been already discussed in the literature so that 
we do not present any further analysis on them. 
The $N_f$ oscillations in density profiles and 
momentum distribution function (MDF), and other mentioned 
characteristics of the 1D trapped system without the lattice 
\cite{vignolo1,minguzzi} are easily obtained in this case.

A qualitative difference between the cases of the trap
with and without a lattice starts for levels in region B. 
Once the degeneracy appears in the spectrum, 
the corresponding eigenfunctions of the degenerate levels start having 
zero weight in the middle of the trap, and for higher levels the regions 
over which the weight is zero increases. As an example, we show in 
Fig.\ \ref{Orbitals}(b) two normalized eigenfunctions belonging 
to region B in Fig.\ \ref{ELV_I0}. The cases depicted correspond to 
the normalized eigenfunctions 5000 (that is only different from zero 
for negative values of $x$) and 5001 (only different from zero for 
positive $x$), for the same parameters of Fig.\ \ref{ELV_I0} (in 
principle a lineal combination of these two eigenfunctions could have 
been the solution since the level is degenerated). Hence, particles in 
these states are confined to a fraction of the trap, showing that the 
combination of both a confining and a periodic potential lead to 
features not present either in the purely confined case without a 
lattice or in the case of a purely periodic potential. Furthermore, 
since we are dealing with a noninteracting case, such features are 
common to both fermions and bosons. However, in the case of fermions, 
it is easy to understand the reason for such effects, as discussed next.

\begin{figure}[h]
\includegraphics[width=0.49\textwidth,height=0.24\textwidth]
{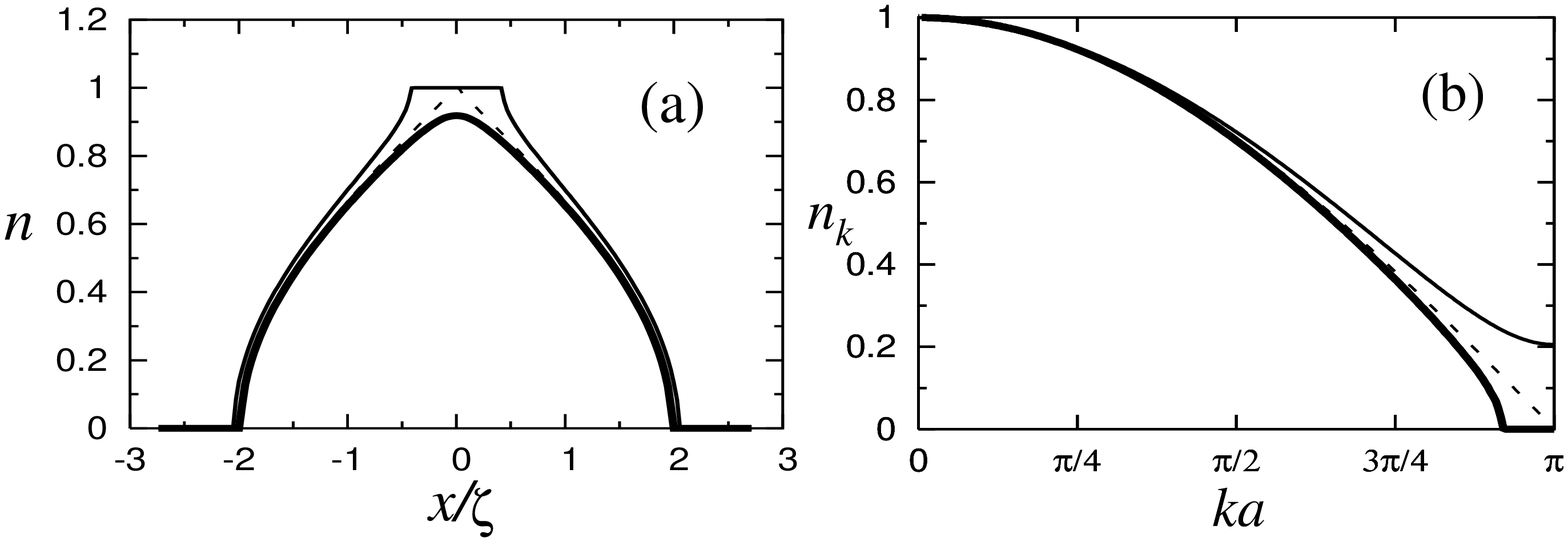}
\caption{Density profiles (a) and normalized MDF (b) for 
$N_f=4500$ (thick continuous line), $N_f=4651$ (dashed line), 
and $N_f=5001$ (think continuous line) 
for a system with $N=10000$ and $V_2a^2=3 \times 10^{-7}t$.
In (a) the positions are given in units of the characteristic length 
$\zeta$, and in (b) the momentum is normalized 
by the lattice constant $a$.}
\label{PerfilsV_I0}
\end{figure}

Figure \ref{PerfilsV_I0}(a) shows density profiles of fermions 
when the number of particles in the trap is increased. 
In one case ($N_f=4500$) the Fermi energy lies just 
below the level marked with an arrow in Fig.\ \ref{ELV_I0}. 
A second curve ($N_f=4651$) corresponds
to the case where the central site reaches a density $n=1$,
and in the other case ($N_f=5001$), the Fermi 
energy lies at the value corresponding to the levels depicted in 
Fig.\ \ref{Orbitals}(b). The positions 
in the trap are normalized in terms of the characteristic length 
for a trapped system when a lattice is present, which is given by 
\cite{rigol03_1,rigol03_2}
\begin{equation}
\label{zeta}
\zeta=\left(V_{2}/t \right)^{-1/2}.
\end{equation}
When the Fermi energy approaches the level where degeneracy sets in,
the density of the system approaches $n=1$ in the middle of the trap, 
and at the filling point where the degeneracy appears in the spectrum, 
the density in the middle of the trap is equal to one, so that 
an insulating region appears in the middle of the system. 
Increasing the filling of the system increases the region over which this 
insulator extends. Hence, due to Pauli principle, the eigenfunctions 
of such levels cannot extend over the insulating region,  
and for the same reason, the region over which the weight is zero 
increases for higher levels. The local insulator with $n=1$ 
has zero variance of the density and, it is incompressible, 
a property that could be tested experimentally by using a local probe.

However, since the system we are considering 
is a noninteracting one, the confined states discussed above should 
be also present in the case of bosons, where the argument about the 
filling would not be valid anymore. It is therefore desirable to also 
understand such features from a single particle perspective \cite{pfau}.
We first notice that the point at which degeneracy appears 
is at an energy $4t$ above the lowest level ($E_0$) 
[see Fig.\ \ref{ELV_I0}(a)], corresponding to the bandwidth for the 
periodic potential. Such an energy is reached when the Bragg condition 
is fulfilled and in the case of the tight-binding system we are considering, 
when all the available states are exhausted. Let us next consider the case 
depicted in Fig.\ \ref{Orbitals}(b). There, the energy level corresponding 
to the wave functions is $E_{5001}-E_0 = 4.2176 t$, that is to a good 
approximation $4t + V_2 x_1^2$ for $V_2a^2=3 \times 10^{-7}t$ and $x_1=697a$
the inner point where the wave functions drop to a value $\sim 10^{-5}$. 
Therefore, the inner turning point corresponds to the Bragg condition, 
whereas for the outer turning point ($x_2=3770a$, again for the same drop 
of the wave function), we have that $E_{5001}-E_0 \simeq V_2 x_2^2$, 
i.e., the classical turning point corresponding to the harmonic 
potential, as expected for such a high level. Hence, Bragg scattering as 
in the well known Bloch oscillations \cite{zener34}, and the trapping 
potential combine to produce the confinement discussed here.  

Further confirmation of the argument above can be obtained by
considering the MDF, a quantity
also accessible in time of flight experiments \cite{greiner}. 
Due to the presence of a lattice, it is a periodic function in the 
reciprocal lattice \cite{kaskurnikov} and it is symmetric with respect 
to $k=0$, so that we study it in the first Brillouin zone 
in the region [$0,\ \pi/a$]. In addition we normalize the 
MDF to unity at $k=0$ ($n_{k=0}=1$). For the fermionic case, it can 
be seen that it always has a region with $n_k=0$ if the insulating phase is 
not present in the trap, and this region disappears as soon as 
the insulator appears in the middle of the system. More precisely,
Fig.\ \ref{PerfilsV_I0}(b) shows that at the filling when the site in
the middle reaches $n=1$, also the momentum $k=\pi/a$ is reached, such that 
the Bragg condition is fulfilled for the first time, confirming the 
discussion above. When further sites reach a density $n=1$, $n_{k=\pi/a}$
increases accordingly. Then the formation of the local insulator in the 
system can be tested experimentally observing the occupation of the 
states with momenta $k=\pm\pi/a$.

\begin{figure}[h]
\includegraphics[width=0.49\textwidth,height=0.23\textwidth]
{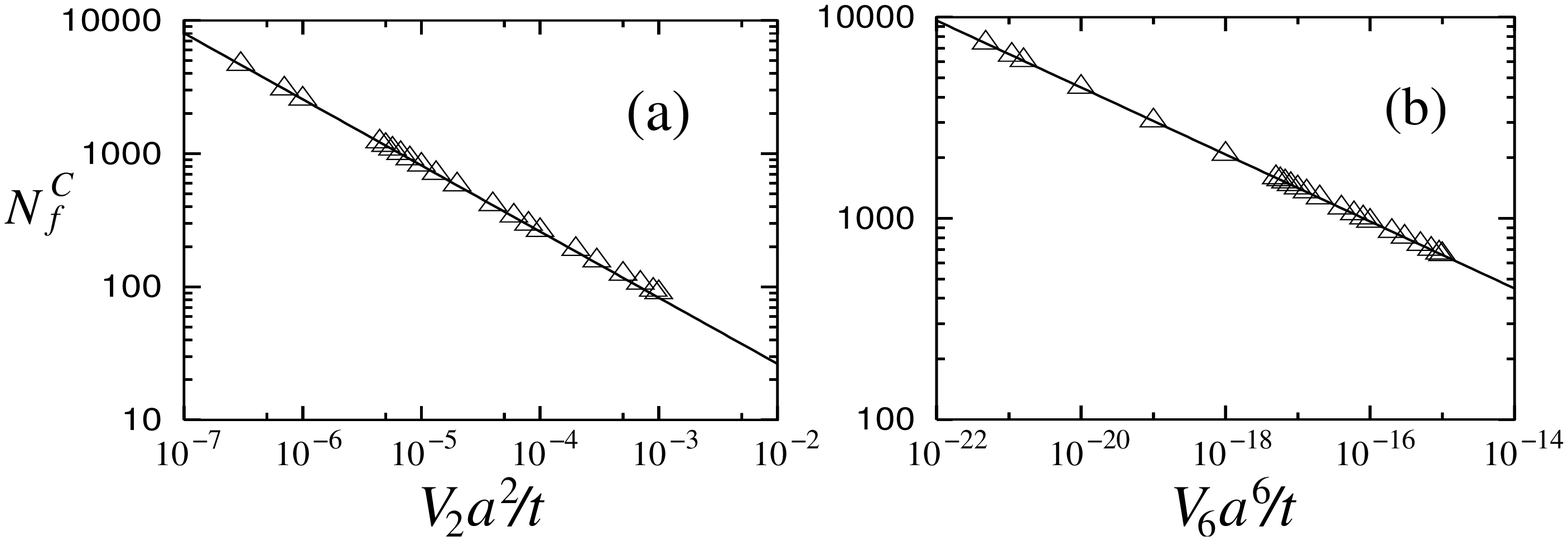}
\caption{Total filling in the trap needed for the formation of 
the insulator as a function of the curvature of the confining potential. 
(a) Harmonic potential. (b) Potential with a power $\alpha=6$.}
\label{ScalingV_I0}
\end{figure}

Since for different systems sizes and number of particles, potentials 
with different curvatures have to be considered, it is important to 
determine the filling $N_f^C$ at which the insulator appears in the 
middle of the trap as a function of the curvature of the harmonic 
confining potential. This question was already answered for the 
interacting case (Hubbard model) in Refs.\ \cite{rigol03_1,rigol03_2} 
where we determined the phase diagram. There we showed that if 
a dimensionless characteristic density $\tilde{\rho}$ is defined 
as $\tilde{\rho}=N_fa/\zeta$, then its value when the insulating regions 
(Mott insulating and band insulating in the interacting case) appear in the
system is always constant for any value of $V_2/t$ at a given value 
of $U/t$ (within error bars there), so that $N^C_f \sim \zeta/a$. 
However, in Refs.\ \cite{rigol03_1,rigol03_2} we were able to check this 
only up to 150 lattice sites and fillings up to the same order, 
whereas here we extend those results to much larger systems. 
In Fig.\ \ref{ScalingV_I0}(a) we show in a log-log scale how $N^C_f$ 
depends on $V_2/t$ over three decades on the total filling. 
In our fit the slope of the curve is $-0.500$ (with 0.04 percent of error), 
as expected on the basis of Eq.\ (\ref{zeta}). The critical characteristic 
density ${\tilde \rho}_C = N_f^C a\left(V_2 /t\right)^{1/2}$ at which the 
insulating region appears is ${\tilde \rho}_C = {\rm e}^\beta$,
with $\beta = 0.986$ (with 0.3 percent of error), 
which is curiously rather close to the basis of the natural logarithms. 

For systems with other powers for the confining potential it 
is only needed to define the appropriate dimensionless characteristic density
$\tilde{\rho}=N_fa\left(V_{\alpha}/t\right)^{1/\alpha}$, and determine 
its value at the point where the insulator appears. In 
Fig.\ \ref{ScalingV_I0}(b) we show in another log-log plot how $N_f^C$ 
depends on the curvature of a confining potential with power six ($V_6/t$). 
As anticipated, we obtain that the slope of the curve is $1/6$ 
(with 0.01 percent of error) in this case and the characteristic density 
for the formation of the insulator is ${\tilde \rho}_C=2.09$. Finally, 
we should mention that it was already shown in Ref.\ \cite{rigol03_2} 
that keeping constant the characteristic density but changing the 
curvature of the confining potential and the total filling in the trap,
the density profiles as a function of the normalized coordinate and 
the normalized MDF remain unchanged.

\begin{figure}[h]
\includegraphics[width=0.43\textwidth,height=0.47\textwidth]
{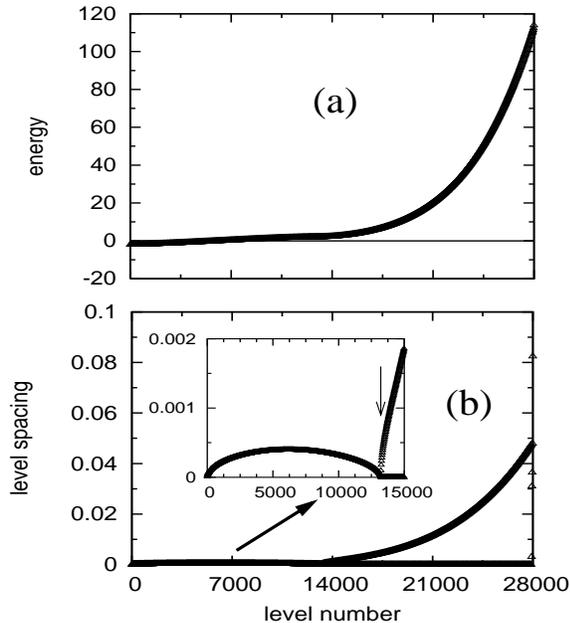}
\caption{Single particle spectrum (a) and level spacing (b) 
for a system with a confining potential with power $\alpha=6$, 
$N=28000$, and $V_6a^6=3 \times 10^{-7}t$. 
Energies are measured in units of $t$. Expanded view in (b) shows 
the first part of the level spacing. An arrow was introduced 
in the inset for signaling the level at which the degeneracy appears 
in the spectrum.}
\label{EL6V_I0}
\end{figure}

In general, for arbitrary confining potentials the same features 
discussed previously for the harmonic case are valid. The
spectrum and level spacing behave in a different way
depending on the power of the confining potential, but always at a certain
level number degeneracy appears in the single particle
spectrum and it corresponds to the formation of an
insulator in the middle of the system for the corresponding filling. 
In Fig.\ \ref{EL6V_I0} we show the single particle spectrum 
[Fig.\ \ref{EL6V_I0}(a)] and the corresponding level spacing [Fig.\ 
\ref{EL6V_I0}(b)] for a confining potential with power $\alpha=6$, 
where the features mentioned previously are evident. 
The arrow in the inset of Fig.\ \ref{EL6V_I0} shows 
the level at which degeneracy sets in, very much in the 
same way as in the harmonic case.

We close this section by considering the pair distribution function.
This quantity not only reflects the consequences of 
Pauli's exclusion principle but clearly characterizes the insulating region.  
In the presence of a lattice the pair distribution function can be 
written as
\begin{equation}
\label{paireq} P_{ij}=\langle n_i \rangle\langle n_j \rangle -
\rho_{ij}^2,
\end{equation}
where $\rho_{ij}=\langle c_i^\dag c_j \rangle$ is the fermionic  
one-particle density matrix.

\begin{figure}[h]
\includegraphics[width=0.36\textwidth,height=0.65\textwidth]
{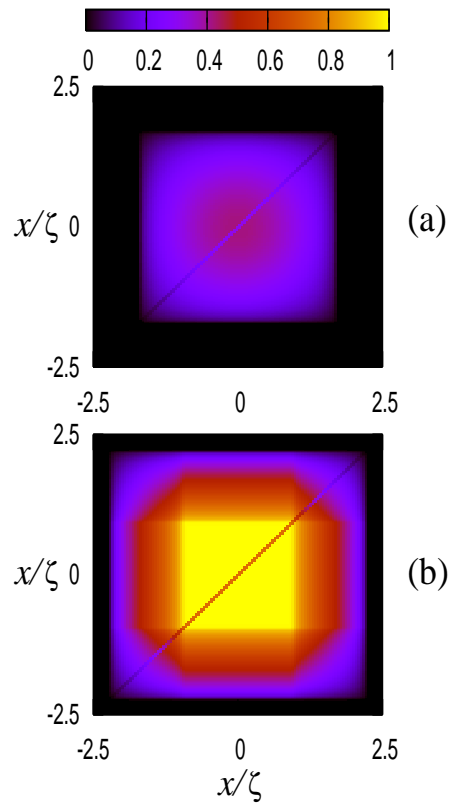}
\caption{(Color online) Intensity plots of the pair distribution function 
as a function of the normalized position for (a) $N_f=300$ 
(only a metal is present in the trap) and (b) $N_f=600$ 
(a insulating region is formed in the middle of the trap). 
The number of lattice sites is $N=1000$ and $V_2a^2=3 \times 10^{-5}t$.}
\label{Pair}
\end{figure}

In Fig.\ \ref{Pair} we show as intensity plots the pair 
distribution function for systems with $N=1000$ lattice sites 
and $V_2a^2=3 \times 10^{-5}t$. Figure \ref{Pair}(a) corresponds 
to the case with $N_f=300$ fermions, where the systems is completely 
metallic, whereas Fig.\ \ref{Pair}(b) corresponds to $N_f=600$ 
fermions, an insulating region appears in the middle of the trap.
Apart from the depression along the diagonal that reveals the 
consequences of Pauli's exclusion principle, a clear distinction 
between the purely metallic case and the one with an insulating region 
can be seen. Inside the insulating region, the density matrix becomes 
diagonal, such that $P_{ij}=1$ for $i\ne j$ and $P_{ii}=0$.

\section{Oscillations of fermions in a 1D lattice}

Recent experiments have realized single species noninteracting fermions 
in 1D optical lattices \cite{modugno03,ott04,pezze04}. Transport studies 
in such systems revealed that under certain conditions a sudden displacement 
of the trap center is followed by oscillations of the c.m. of the fermionic 
cloud in one side of the trap. This is in contrast to the system without 
the lattice where the c.m. oscillates, as expected, around the potential 
minimum \cite{modugno03,ott04,pezze04}. Although the experimental system is 
not a true 1D system, due to the strong transversal confinement the relevant 
motion of the particles occurs in the longitudinal direction. Hence,  
in order to qualitatively understand the observed behavior one can 
analyze the ideal 1D case. Given the results discussed in the previous 
section one expects the displaced oscillation of the c.m. to appear when, 
due to the initial displacement of the trap, particles that where 
located in region A of the spectrum in Fig.\ \ref{ELV_I0}(b) are moved 
into region B so that Bragg conditions are fulfilled. Then the 
particles get trapped in one side of the system 
[Fig.\ \ref{Orbitals}(b)].

\begin{figure}[h]
\includegraphics[width=0.47\textwidth,height=0.29\textwidth]
{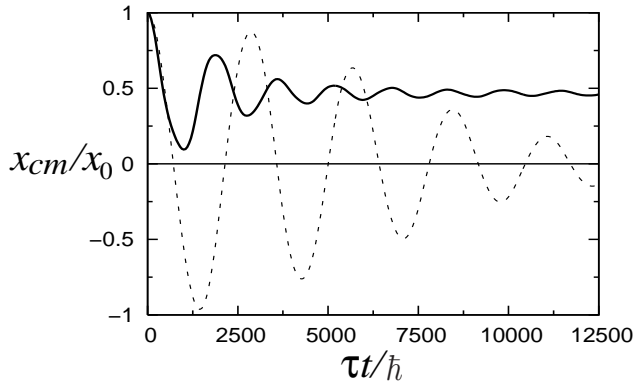}
\caption{Evolution of the c.m. ($x_{cm}$) of 1000 confined fermions 
when the center of the trap is suddenly displaced 200 lattice sites ($x_0$), 
for $V_2a^2=2\times 10^{-06}t$ (dashed line), 
and $V_2a^2=6\times 10^{-06}t$ (continuous line).} 
\label{CM}
\end{figure}

\begin{figure}[h]
\includegraphics[width=0.49\textwidth,height=0.23\textwidth]
{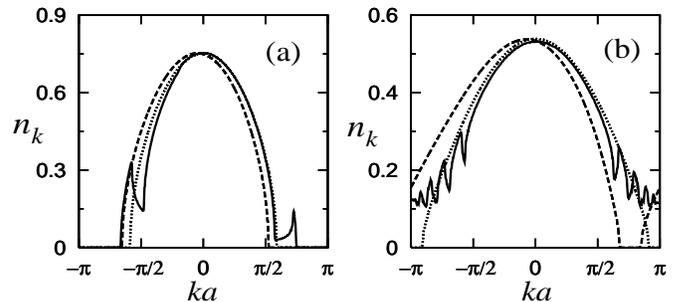}
\caption{MDF of 1000 trapped fermions at three different times 
after displacing the trap 200 lattice sites, for 
$V_2a^2=2\times 10^{-06}t$ (a), and $V_2a^2=6\times 10^{-06}t$ (b).
The times are $\tau=0$ (dotted line), 
$\tau=250\hbar/t$ (dashed line), 
and $\tau=12500\hbar/t$ (continuous line). }
\label{PerfilKTIME}
\end{figure}

Figure \ref{CM} shows exact results obtained for the c.m. dynamics of 
1000 fermions in a trap with $N=3000$ when its center is 
suddenly displaced 200 lattice sites. $\tau$ denotes the real time 
variable. The relation between the confining potential ($V_2$) and the 
hopping parameter ($t$) is increased in order to fulfill the Bragg 
conditions. This is equivalent in experiments to increase the curvature 
of the confining potential keeping constant the depth of the lattice, 
which leads to an increase of the frequency of the oscillation as shown in 
Fig.\ \ref{CM}. It is also equivalent to increase the depth of the lattice 
keeping the confining potential constant, but then our plots in Fig.\ \ref{CM} 
should be interpreted with care since there we normalize the time variable 
by the hopping parameter, which changes in the latter case.
In Fig.\ \ref{CM} (dashed line) we show results for 
the case where the c.m. of the cloud oscillates around the minimum of energy 
of the trap since no Bragg conditions are fulfilled. This can be seen 
in the MDF [Fig.\ \ref{PerfilKTIME}(a)] where at any time no particles 
have $k=\pm \pi/a$. Figure \ref{CM} also shows that a damping of the 
oscillation of the c.m. occurs. This is due to the nontrivial dispersion 
relation in a lattice $\epsilon_k=-2t \cos ka$, which makes the frequency 
of oscillation of the particles dependent on their energies, leading to 
dephasing. In order to reduce the damping, fermions should populate after 
the initial displacement only levels with energies close to the bottom 
of the band in a lattice, so that the quadratic approximation is valid for 
$\epsilon_k$. (Notice that this is not generally fullfiled even if the 
initial displacement is small.) 

Increasing the relation $V_2/t$ makes that some particles start to fulfill 
the Bragg conditions so that the center of oscillations of the cloud depart 
from the middle of the trap. In Fig.\ \ref{CM} (continuous line) we show 
a case where the c.m. never crosses the center of the trap. The MDF 
corresponding to this case, at three different times, is displayed in 
Fig.\ \ref{PerfilKTIME}(b). There it can be seen that initially ($\tau=0$) 
no Bragg conditions are satisfied in the system, and that some time after 
the initial displacement the Bragg conditions are fulfilled 
($\tau=250\hbar/t$). Finally, we also show the MDF long time after the 
initial displacement of the trap ($\tau=12500\hbar/t$), when the 
oscillations of the c.m. are completely damped and the MDF is approximately 
symmetric around $k=0$.

\section{Doubling the periodicity}

In this section we study the consequences of enlarging the periodicity  
in the lattice. For this purpose we introduce an alternating potential, 
and the Hamiltonian of the system can be written as
\begin{eqnarray}
H & = & -t \sum_{i} \left( c^\dagger_{i} c^{}_{i+1} + \text{H.c.}
\right) + V_{\alpha} \sum_{i} \left(x_i \right)^{\alpha}\ 
n_{i } \nonumber \\ & & + V_a  \sum_{i } \left(-1\right)^i  \ n_{i }, 
\label{HubbIon}
\end{eqnarray}
where the last term represents the oscillating potential and
$V_a$ its strength. The purpose of introducing an alternating 
potential in the trapped system is twofold. For fermionic systems, 
the increase of the numbers of sites per unit cell leads to the 
possibility of creating insulating states (band insulators in the 
unconfined case) for commensurate fillings. On the other hand, 
by changing the periodicity, new Bragg conditions are introduced, 
giving the possibility of further control on the confinement discussed 
in the previous sections.

\begin{figure}[h]
\includegraphics[width=0.98\textwidth,height=0.32\textwidth]
{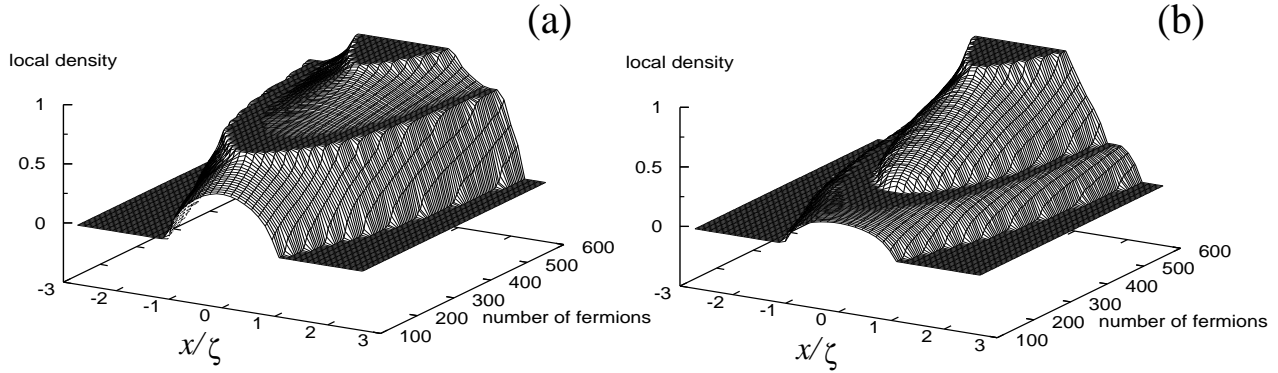}
\caption{Evolution of the local density in a harmonic trap 
as a function of the position and increasing total number of
fermions when an alternating potential $V_a=0.5 t$ is present. 
(a) Odd sites. (b) Even sites. The system has 1000 lattice sites 
and $V_2a^2=3\times 10^{-5}t$.}
\label{Perfil3D}
\end{figure}

Figure \ref{Perfil3D} shows how the density profiles evolve
in a harmonic trap when the total filling is increased. Since the density 
oscillates due to the alternating potential, we made two 
different plots for the odd [negative value of the alternating potential, 
Fig.\ \ref{Perfil3D}(a)] and even [positive value of the 
alternating potential, Fig.\ \ref{Perfil3D}(b)] sites. Each of the plots 
in Fig.\ \ref{Perfil3D} is very similar to the evolution of the density 
profiles already shown for the trapped Hubbard model 
\cite{rigol03_1,rigol03_2}. The only difference is that in 
Figs.\ \ref{Perfil3D}(a) and \ref{Perfil3D}(b) the plateaus
with $n\ne 1$ have densities different between themselves and 
different from $n=0.5$, which would be the density of one component of 
the spin polarized fermions in the Mott insulating phase of the Hubbard 
model. In the flat regions of Fig.\ \ref{Perfil3D}, both even and odd sites 
have the same densities than the corresponding sites in the periodic 
case at half filling for the same value of the alternating potential, 
so that it is expected that they correspond to local insulating phases.

\newpage

\onecolumngrid

\twocolumngrid

\begin{figure}[h]
\includegraphics[width=0.43\textwidth,height=0.44\textwidth]
{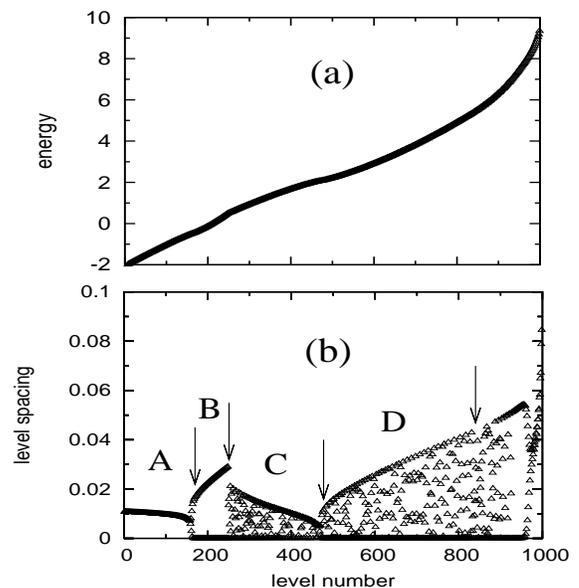}
\caption{Single particle spectrum (a) and level spacing (b) 
for a system with an alternating potential $V_a=0.5 t$ and 
with $N=1000$ and $V_2a^2=3 \times 10^{-5}t$.
Energies are measured in units of $t$.
For the explanation of the regions between the arrows, see text.}
\label{ELV_I05}
\end{figure}

In Figs.\ \ref{ELV_I05}(a) and  \ref{ELV_I05}(b) we show the single 
particle spectrum and the level spacing respectively for the same 
parameters of Fig.\ \ref{Perfil3D}. Although in this case the level 
spacing exhibits a more complicated structure, an immediate identification 
between the regions signaled in Fig.\ \ref{ELV_I05}(b) between arrows 
and different fillings in Fig.\ \ref{Perfil3D} can be done. (A) 
corresponds to the fillings in Fig.\ \ref{Perfil3D} where only a 
metallic phase appears in the trap, (B) to the fillings 
where the first plateau is present in Fig.\ \ref{Perfil3D}, (C) 
to the fillings where a metallic phase develops in the middle of 
the trap and it is surrounded by insulating regions, and (D) to 
the fillings in Fig.\ \ref{Perfil3D} where the insulator with $n=1$ 
appears in the center of the system. The region after the last arrow 
in Fig.\ \ref{ELV_I05}(b) corresponds to deconfined states. 
Notice that the level spacing in regions (C) and (D) shows a behavior 
that was not present in Fig.\ \ref{ELV_I0}.

\begin{figure}[h]
\includegraphics[width=0.48\textwidth,height=0.23\textwidth]
{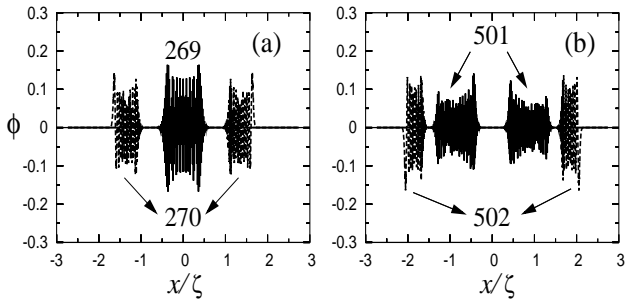}
\caption{Eigenfunctions for a trapped system with an alternating 
potential ($V_a=0.5t$) for $N=1000$ and $V_2a^2=3 \times 10^{-5}t$. The 
eigenfunctions correspond to the levels: 269 (a) (continuous line), 
270 (a) (dashed line), 501 (b) (continuous line), and 502 (b) dashed 
line.}
\label{OrbitalsV_I05}
\end{figure}

In order to understand the complex behavior of the level spacing 
we study, as in the previous section, the eigenfunctions of the 
system shown in Fig.\ \ref{ELV_I05}. The eigenfunctions 
corresponding to region A in Fig.\ \ref{ELV_I0}(b) 
behave as expected for a metallic phase, where the combination of 
the alternating and confining potentials generates a different 
modulation than the one studied in Sec.\ II, but without qualitative 
differences. In the second region of the spectrum (region B) the 
eigenfunctions have zero weight in the middle of the trap, exactly like 
in the insulator discussed in Sec.\ II. In region C there is, 
as pointed out above, a new feature since in this case it is possible to 
obtain a metallic region surrounded by an insulating one. This is reflected 
by the eigenfunctions shown in Fig.\ \ref{OrbitalsV_I05}(a), where one of the 
eigenfunctions is nonzero only inside the local insulating phase (continuous 
line), and the other is nonzero only outside the insulating phase (dashed 
line), the energy levels associated with the latter ones are degenerated. 
For the region D the situation is similar but in this case the system is 
divided in four parts because of the existence of the insulator with $n=1$ 
in the middle of the trap and the insulator between the two metallic phases. 
This implies that all the levels are degenerate in region D, and the 
particles are located either between both insulating regions or outside 
the outermost one, as shown in Fig.\ \ref{OrbitalsV_I05}(b).

As in the previous section, the spectral features discussed here are 
equally valid for fermions as well as for bosons. Up to now we discussed 
the ``slicing'' of the systems only in terms of fermions and based on the 
appearance of insulating regions along the system. As before, it would be 
also here desirable to understand the appearance of forbidden regions in
space in terms of a single particle picture. We show now that
with the introduction of new Bragg conditions, due to the altered 
periodicity, the ``slicing'' of the system can be explained in an analogous 
way as in the previous section. In the unconfined case, the doubling of the 
periodicity creates new Bragg conditions at $k=\pm \pi/2a$, such that an 
energy gap $2 V_a$ appears. Figure \ref{ELV_I05}(a) shows that in the 
confined case the spectrum is continuous (in the sense that the level 
spacing is much smaller than $2 V_a$), so that the imprint of the gap 
can be seen only in the local density of states
\begin{equation}
N_i (\omega) = \frac{1}{\pi} \mbox{Im} G_{ii} (\omega) \; ,
\end{equation}
where $G_{ij} (\omega)$ is the one-particle Green's function 
\cite{mahan}, which in this case can be easily computed.

\begin{figure}[h]
\includegraphics[width=0.42\textwidth,height=0.28\textwidth]{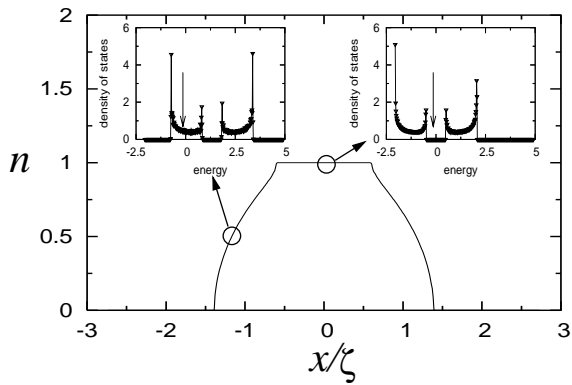}
\caption{Density profile per unit cell (now containing two contiguous 
lattice sites) for a trap with $N=10000$, $V_2a^2=3 \times 10^{-7}t$, and 
$N_f=2000$. The insets show the density of states per unit cell for two 
points in the profile. The arrows in the insets signal the Fermi energy 
for the selected filling.}
\label{DensStates}
\end{figure}
The insets in Fig.\ \ref{DensStates} show the density of states per unit 
cell (now containing two lattice points) for two different 
positions along the density profile. The downward arrows in each inset 
corresponds to the location of the Fermi energy. The inset at the left 
corresponds to a situation where the Fermi energy goes through the lowest 
band, whereas the inset at the right belongs to sites in the middle of an 
insulating region. As expected, in this latter case, the Fermi energy 
lies inside the gap. The size of the gap is to a high degree of accuracy 
$2 V_a$ for the site in the middle of the trap, but slightly less on the 
sides. Therefore, again the same arguments as before can be used, but 
instead of $4t$, the width for each band is given by 
$\sqrt{4 t^2 +V_a^2} - V_a \simeq 1.56 t$ in our case. Without repeating
the detailed discussion in the previous section, we can understand 
the confinement in Fig.\ \ref{OrbitalsV_I05}(a) as follows. Level 269
has an energy that for sites in the middle of the trap falls in
the middle of the upper band, while for level 270 (the same value 
of energy), passes through the lowest band. In fact, the density of 
states shown in Fig.\ \ref{DensStates} can be viewed
as approximately shifted by $V_2 x_i^2$, counting the sites from the 
middle. Finally, levels in Fig.\ \ref{OrbitalsV_I05}(b) correspond 
to the case where in the middle of the trap they fall beyond the highest 
band, then going outwards, they fall in the middle of the highest band, 
and further outside, they fall in the middle of the lowest band.

\begin{figure}[h]
\includegraphics[width=0.49\textwidth,height=0.67\textwidth]
{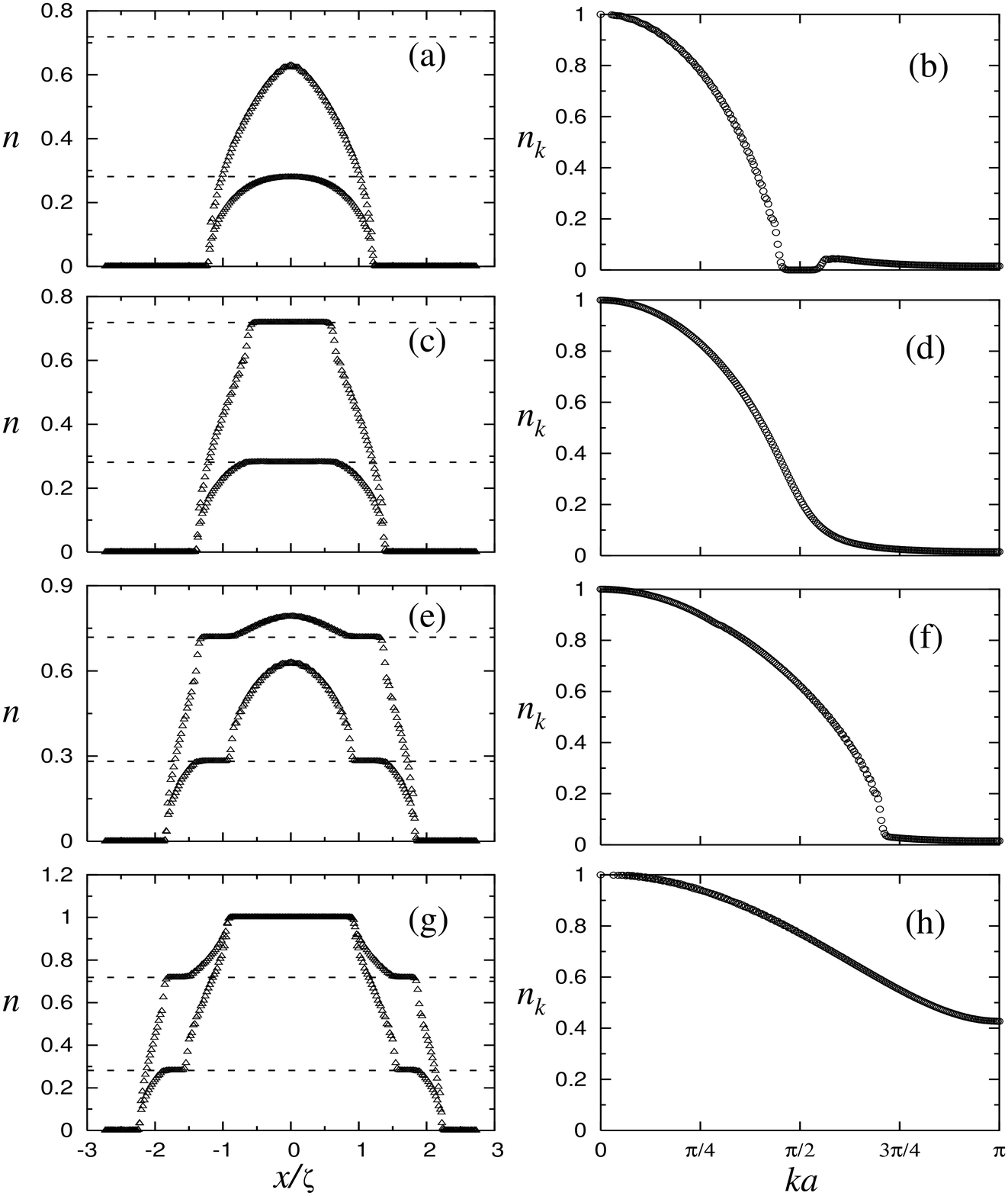}
\caption{Density profiles (left) and their normalized MDF 
(right) for $N_f=150$ (a),(b), 200 (c),(d), 
350 (e),(f), 600 (g),(h) and $N=1000$, $V_a=0.5t$, 
$V_2a^2=3 \times 10^{-5}t$.}
\label{Perfiles1000}
\end{figure}
Again, as in Sec.\ II, one can follow the same reasoning 
by considering the MDF. Due to the new periodicity it
displays new features, associated with the fact that increasing
the periodicity the Brillouin zone is decreased, and in the present 
case a second Brillouin zone is visible. 
In Fig.\ \ref{Perfiles1000} we show the density profiles 
(left) that characterize the four different situations 
present in Fig.\ \ref{Perfil3D}. They correspond to fillings of 
the trap in the four regions of the single particle spectrum 
discussed previously in Fig.\ \ref{ELV_I05}. Notice that in the figures 
we included all the odd and even points in the density profiles. 
We plotted as horizontal dashed lines the values of the densities 
in the band insulating phase of the periodic system for the odd and 
even sites (from top to bottom respectively), so that 
it can be seen where are located the local insulating phases in the 
trap. The corresponding normalized MDF are presented in 
Fig.\ \ref{Perfiles1000} (right). 

In Figs.\ \ref{Perfiles1000}(a) and \ref{Perfiles1000}(b) it is possible 
to see that when only the metallic phase is present in the trap, 
in the MDF an additional structure appears after $\pi/2$, 
corresponding to the contribution from the second Brillouin zone.
When a first insulating phase is reached, by coming to the top of
the lowest band, $k=\pi/2a$ is reached, and increasing the fillings 
of the system beyond that point, the dip around $k=\pi/2a$ disappears 
[Figs.\ \ref{Perfiles1000}(c) and \ref{Perfiles1000}(d)].
On adding more particles to the system a metallic phase appears 
inside the insulating plateau, [Figs.\ \ref{Perfiles1000}(e) and 
\ref{Perfiles1000}(f)]. When this metallic phase widens, decreasing the 
size of the insulating phase, $n_k$ starts to be similar to the $n_k$ 
of the pure metallic phase in the system without the alternating potential.
Increasing even further the filling of the system, when the trivial 
insulator ($n=1$) appears in the center of the trap, the tail with very 
small values of $n_k$ disappears (like in the system without the 
alternating potential the region with $n_k$ zero also disappears 
[Fig.\ \ref{PerfilsV_I0}(b)] and the further increase of the filling in the 
system makes $n_k$ flatter [Figs.\ \ref{Perfiles1000}(h) and 
\ref{Perfiles1000}(i)]. 

Up to this point, several quantities, like density profile, pair
distribution function, or local density of states 
were taken as evidence for the existence of an insulating phase, but a 
quantitative criterion in the sense of an order parameter 
to characterize the phases was not given. As shown already
in the case of the Hubbard model \cite{rigol03_1,rigol03_2}, it
is possible to define a local compressibility:
\begin{equation}
\label{localc} \kappa_i^\ell = \sum_{\mid j \mid \leq \, \ell (V_a)}
\chi_{i,i+j} \ ,
\end{equation}
where
\begin{equation}
\chi _{i,j}=\left\langle n_{i}n_{j}\right\rangle -\left\langle
n_{i} \right\rangle \left\langle n_{j}\right\rangle
\end{equation}
is the density-density correlation function, and $\ell (V_a) \simeq
b\, \xi (V_a)$, with $\xi (V_a)$ the correlation length of
$\chi_{i,j}$ in the periodic system at half-filling for the
given value of $V_a$. As a consequence of the band gap opened in
the band insulating phase at half filling in the periodic
system, density-density correlations decay exponentially and
there $\xi (V_a)$ can be determined. The parameter $b$ is 
considered $b\sim 10$ (see discussion in Ref.\ \cite{rigol03_2}). 
When this definition is applied to the different
fillings of Fig.\ \ref{Perfil3D} the local compressibility is zero 
in the insulating regions and nonzero in the metallic phases. 
The local quantum critical behavior found in Ref.\ \cite{rigol03_1} 
at the transition between the metallic and Mott insulating phase 
is not present here since there are no interactions between the 
particles that could generate quantum criticality.

\begin{figure}[h]
\includegraphics[width=0.4\textwidth,height=0.26\textwidth]
{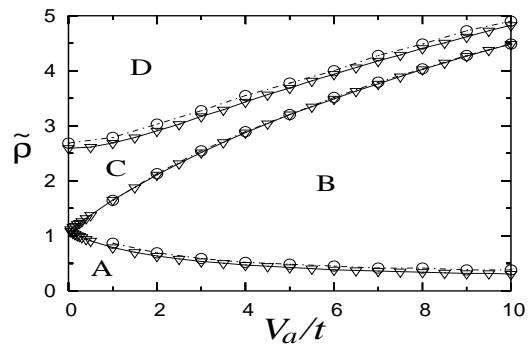}
\caption{Phase diagram for systems with $V_2a^2=3 \times 10^{-5}t$ 
($\bigtriangledown$) and $V_2a^2=3 \times 10^{-4}t$ 
($\bigcirc$). The different phases are explained 
in the text.}
\label{DFS_ScalingF}
\end{figure}

Finally we analyze the phase diagram for these systems.
It can be generically described by the characteristic density 
$\tilde{\rho}$, like the Hubbard model and the noninteracting 
case in Sec.\ II. In Fig.\ \ref{DFS_ScalingF} we show two phase 
diagrams for two different values of the curvature of the confining 
potential, $V_2a^2=3\times 10^{-5}t$ and $V_2a^2=3\times 10^{-4}t$.
There it can be seen that although there is one order of magnitude 
between the curvatures of the confining potentials, the phase diagrams 
are one on top of the other, the small differences are only due to 
the finite number of particles which make the changes in $\tilde \rho$ 
discrete. Therefore, the characteristic density 
allows us to compare systems with different curvatures of the 
confining potential, number of particles and sizes. In addition 
we checked that keeping the characteristic density constant 
for a given value of $V_a$, the density profiles as a function of 
the normalized coordinates and the normalized MDF  
do not change when the number of particles or the curvature of the 
confining potential are changed in the system, as we already pointed out 
for the case without alternating potential (Sec.\ II). 

The different phases present in Fig.\ \ref{DFS_ScalingF} are (A)
a pure metallic phase, (B) an insulator in the middle of the trap 
surrounded by a metallic phase, (C) a metallic intrusion in the middle 
of the insulator, (D) an insulator with $n=1$ in the center of the trap 
surrounded by a metal, an insulator and the always present external 
metallic phase. For very small values of the alternating potential ($V_a$), 
phase B is not present, and the insulator surrounding the metallic phase 
in the center of the trap disappears leaving a full metallic phase at 
the very beginning of phase C. Similarly, the insulator with $n=1$ is 
surrounded only by a metallic phase (at the very beginning of phase D). 
However, these regions are very small in the phase diagram and we did not 
include them. At $V_a=0$ the results of Sec.\ II are recovered since 
up to $\tilde{\rho}=2.68$ the system is a pure metal and for higher 
characteristic densities there is an insulator with $n=1$ surrounded by 
metallic phases.

There is one important difference between the phase diagram in Fig.\ 
\ref{DFS_ScalingF} and the one of the trapped Hubbard model 
\cite{rigol03_1,rigol03_2}. In Fig.\ \ref{DFS_ScalingF} the boundary 
between regions A and B changes appreciably when the value on 
the alternating potential is increased while in the Hubbard model 
case (for the values of $U$ that we simulated) it was found independent 
of the value of $U$. This is possibly due to the fact that
for the alternating potential, increasing $V_a$ changes 
the local densities of the insulating phase while in the 
local Mott insulating phase the density is always constant 
independently of the value of $U$.

\section{The 2D system}

In this section we extend to 2D the results obtained in
previous sections for the 1D case. The Hamiltonian in this
case can be written as 
\begin{equation}
\label{Ham2D} H  =  -t \sum_{\langle i,j \rangle} 
\left( c^\dagger_{i}c^{}_{j} + \text{H.c.} \right)
+ \sum_{i} \left( V_{\alpha_x,x}\ x_i^{\alpha_x} +
V_{\alpha_y,y}y_i^{\alpha_y} \right) n_{i},
\end{equation}
where ($x_i,y_i$) are the coordinates of the site $i$, and 
$\langle i,j \rangle$ refers to nearest neighbors. 
The last term in Eq.\ (\ref{Ham2D}) allows to consider different 
strengths $V_{\alpha_x,x},\ V_{\alpha_y,y}$ and powers 
$\alpha_x,\ \alpha_y$ of the confining potential in the $x$, $y$ 
directions. We call in what follows $N_x$ and $N_y$ 
the number of lattice sites in the $x$ and $y$ directions, 
respectively. 

\begin{figure}[h]
\includegraphics[width=0.43\textwidth,height=0.43\textwidth]
{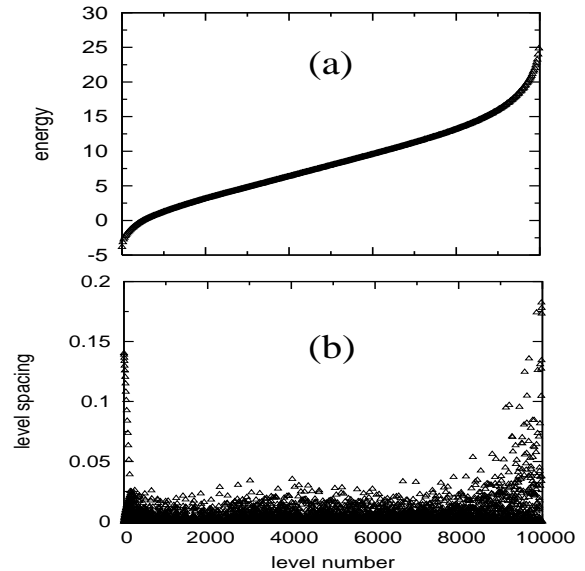}
\caption{Single particle spectrum (a) and level spacing (b) for a 
system with $N_x=N_y=100$, and $V_{2,x}a^2=V_{2,y}a^2=
5 \times 10^{-3}t$. Energies are measured in units of $t$.}
\label{EL2DV_I0}
\end{figure}

In Fig.\ \ref{EL2DV_I0} we show the single particle spectrum and 
its corresponding level spacing for a system with $N_x=N_y=100$ 
lattice sites confined by a harmonic potential with
$V_{2,x}a^2=V_{2,y}a^2=5 \times 10^{-3}t$. Figure \ref{EL2DV_I0}(b) 
shows that degeneracy sets in at the very beginning of the expectrum, 
and this is because of the symmetries of the square lattice. 
In 2D the formation of the insulator in the middle of the trap 
does not generate additional degeneracies in the system since 
it does not split the trap in independent identical parts. 
Then in contrast to the 1D case no information of its formation 
can be obtained from the level spacing.

\newpage

\onecolumngrid

\begin{figure}[h]
\includegraphics[width=0.94\textwidth,height=0.455\textwidth]
{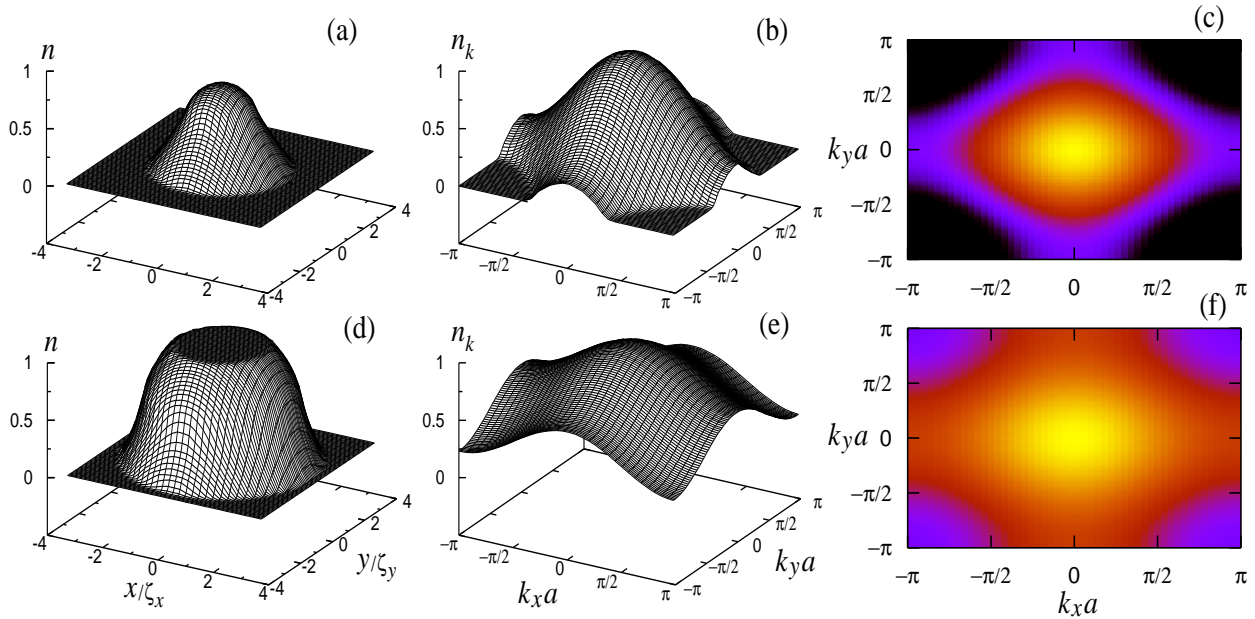}
\caption{(Color online) 2D density profiles (a),(d), 2D normalized MDF 
(b),(e), and intensity plots of the 2D normalized MDF (c),(f), 
for $N_f=1000$ (a)--(c) and $N_f=4000$ (d)--(f) fermions in a system 
with $N_x=N_y=100$ and $V_{2,x}a^2=V_{2,y}a^2=5 \times 10^{-3}t$. 
The color scale in (c) and (f) is the same as in 
Fig.\ \ref{Pair}.}
\label{Perfil100X100}
\end{figure}

\twocolumngrid

Two density profiles and their corresponding normalized MDF 
for $N_f=1000$ and $N_f=4000$, and the same trap
parameters of Fig.\ \ref{EL2DV_I0}, are presented in 
Fig.\ \ref{Perfil100X100}. The $x$ and $y$ coordinates in the trap are 
normalized by the characteristic lengths 
$\zeta_x=\left(V_{2,x}/t \right)^{-1/2}$ 
and $\zeta_y=\left(V_{2,y}/t \right)^{-1/2}$, respectively. 
The density profile in Fig.\ \ref{Perfil100X100}(a) corresponds to a
pure metallic phase in the 2D trap, the corresponding MDF 
[Fig.\ \ref{Perfil100X100}(b)] is smooth and for 
some momenta it is possible to see that $n_k=0$ like in the 1D case.
When the filling of the system is increased, the insulator
appears in the middle of the trap [Fig.\ \ref{Perfil100X100}(d)] and all 
the regions with $n_k=0$, present in the pure metallic phase,
disappear from the MDF [Fig.\ \ref{Perfil100X100}(e)]. 
Figs.\ \ref{Perfil100X100}(c) and \ref{Perfil100X100}(f) show as 
intensity plots the normalized MDF of Figs.\ \ref{Perfil100X100}(b) and 
\ref{Perfil100X100}(e).

In 2D it is possible to define a dimensionless characteristic density as 
$\tilde{\rho}=N_fa^2/\zeta_x\zeta_y$. Also in this case it has always 
the same value when the insulator appears in the middle of 
the system, independently of the values and relations between 
$V_{2x}$ and $V_{2y}$. The density profiles as function of the
normalized coordinates and the MDF
remain unchanged when the characteristic density is kept constant
and the values and relations between $V_{2x}$ and $V_{2y}$ are
changed (in the thermodynamic limit they have the same form 
shown in Fig.\ \ref{Perfil100X100}). This implies that the results 
shown in Fig.\ \ref{Perfil100X100} for a symmetric trap do not 
change for an asymmetric trap with the same characteristic density. 
The value of the characteristic density for the formation 
of the insulator in a harmonic 2D trap is $\tilde{\rho}_C\sim 13.5$. 

\begin{figure}[h]
\includegraphics[width=0.49\textwidth,height=0.86\textwidth]
{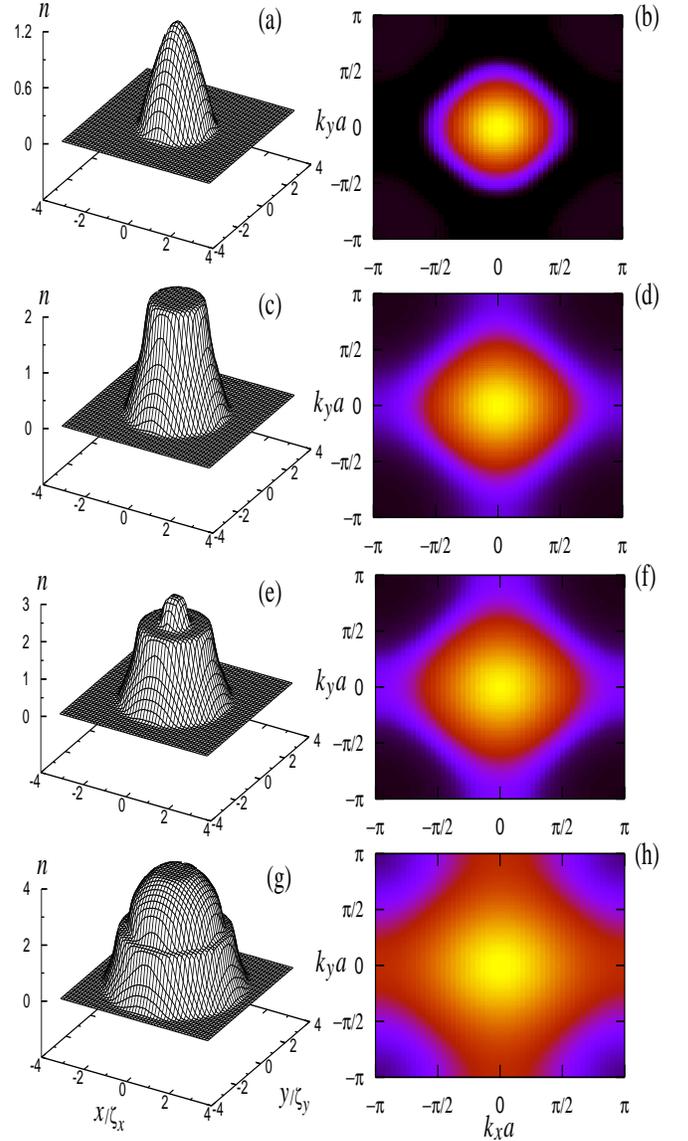}
\caption{(Color online) Density per unit cell (a),(c),(e),(g) 
and intensity plots of the normalized MDF (b),(d),(f),(h) 
profiles for $N_f=200$ (a),(b), 800 (c),(d), 1100 (e),(f), 
and 3000 (h),(i) in a system with $N_x=N_y=100$, 
$V_{2,x}a^2=V_{2,y}a^2=5 \times 10^{-3}t$, and $V_a=t$. 
The color scale in the intensity plots of the normalized MDF is the 
same as in Fig.\ \ref{Pair}.}
\label{Perfil100X100I}
\end{figure}
The addition of the alternating potential leads to results similar to 
those presented in the 1D case. Four density profiles showing the 
possible local phases in the 2D trap, and intensity plots of their 
corresponding MDF are shown in Fig.\ \ref{Perfil100X100I}. 
In the pure metallic case [Figs.\ \ref{Perfil100X100I}(a) and 
\ref{Perfil100X100I}(b)] the additional structure in the MDF 
for $k_x,k_y>\pi/2a$, due to the increase of the periodicity, is present. 
This structure also disappears when the insulator appears in the middle of 
the system [Figs.\ \ref{Perfil100X100I}(c) and \ref{Perfil100X100I}(d)]. 
Increasing the filling a new metallic phase appears in the center of the trap 
[Figs.\ \ref{Perfil100X100I}(c) and \ref{Perfil100X100I}(d)]. For the highest 
filling the insulator with $n=1$ develops in the middle of the trap and the 
MDF becomes flatter with $n_k\neq 0$ everywhere. In a way similar to the 1D 
case there are confined states in the radial direction, i.e., particles are 
confined in rings around the center of the trap, and they can be explained 
in terms of Bragg conditions. The phase diagram of the 2D case is also 
similar to the one in the 1D case and is not discussed here.

\begin{figure}[h]
\includegraphics[width=0.48\textwidth,height=0.50\textwidth]
{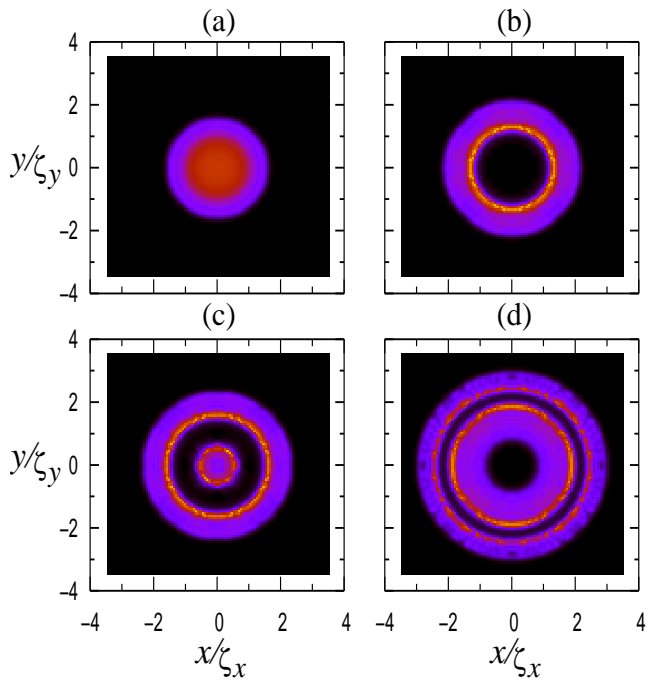}
\caption{(Color online) Intensity plots of the local 
compressibility as a function 
of the normalized coordinates for systems with $N_x=N_y=100$, 
$V_{2,x}a^2=V_{2,y}a^2=5 \times 10^{-3}t$ and $V_a=t$. 
The fillings are the same as that in Fig.\ \ref{Perfil100X100I}
(a) $N_f=200$, (b) $N_f=800$, (c) $N_f=800$, and (d) $N_f=1100$. 
Black color means zero compressibility.}
\label{Comp100X100I}
\end{figure}
We close this section by considering the local compressibility. As it was 
mentioned in the analysis of the 1D case with the alternating potential, this
quantity is zero in the insulating phases (like in the Mott insulating phases 
of the trapped Hubbard model). In the 2D case we extend the definition given 
by Eq.\ (\ref{localc}) to
\begin{equation}
\label{localc2D} \kappa_{i_x i_y}^\ell = 
\sum_{\mid j_x,j_y \mid \leq \, \ell (V_a)}
\chi_{i_x i_y,i_x+j_x i_y+j_y} \ ,
\end{equation}
where
\begin{equation}
\chi _{i_x i_y,j_x j_y}=\left\langle n_{i_x i_y}n_{j_x j_y} 
\right\rangle -\left\langle n_{i_x i_y} 
\right\rangle \left\langle n_{j_x j_y}\right\rangle,
\end{equation}
is the density-density correlation function in 2D and 
$\ell (V_a) \simeq b\, \xi (V_a)$. In this case it is also possible 
to determine $\xi (V_a)$ in the insulating phase
of the 2D periodic case (at half filling) and apply the new
definition to the 2D trap (as for the 1D case $b\sim 10$). 
The results obtained for the same parameters of Fig.\ \ref{Perfil100X100I} 
are presented in Fig.\ \ref{Comp100X100I}. There it can be seen that the 
rings of local insulators in Fig.\ \ref{Perfil100X100I} are represented 
in Fig.\ \ref{Comp100X100I} by rings of incompressible regions (black rings) 
so that this definition works also perfectly in the 2D case, and the 
local compressibility should be a relevant quantity to characterize the 
local Mott insulating phases also in the trapped 2D Hubbard model.

\section{Conclusions}

We performed a detailed analysis of noninteracting systems focusing 
on the consequences of the combination of a confining and a periodic 
potential. It leads to a confinement of particles in a fraction of the 
available system size. This confinement is directly related to the formation 
of insulating regions in the case of fermionic systems. Since the results 
obtained correspond to noninteracting particles they can be also explained 
in a single particle picture due to the realization of Bragg conditions, 
and are also valid for bosons. We have studied the consequences 
of the previous confinement in the nonequilibrium dynamics of trapped 
particles in 1D when the center of the trap is suddenly displaced, 
and confirmed evolution of the center of mass obtained in recent 
experiments. 

The region over which particles are confined in the trap can be controlled 
in various ways. The most obvious one is by changing the strength of the 
confining potential, where the extension of such regions can be regulated. 
Other way is changing the periodicity of the lattice, which leads to 
a different ``slicing'' of the system. The change of the periodicity 
also generates in the fermionic case the possibility of obtaining local 
insulating phases with sizes that can be controlled changing the strength 
of the additional alternating potential. This gives rise to a picture that 
is similar in some aspects to the the Hubbard model analyzed in 
Refs.\ \cite{rigol03_1,rigol03_2}. We have shown that although insulating 
phases appear in this noninteracting case, the gaps that are locally opened 
are not seen in the single particle spectrum. In order to observe 
them it is necessary to study the local density of states. 
The local compressibility defined in Refs.\ \cite{rigol03_1,rigol03_2} 
was also proven to be a genuine local order parameter to characterize the new 
insulating phases since it is always zero there. A scalable phase diagram for 
these systems was also presented. Finally, we considered the two-dimensional 
case and the formation of insulating regions due to the presence of periodic 
potentials. We showed that the local compressibility also characterizes those 
2D regions in an unambiguous way.

\begin{acknowledgments}

We gratefully acknowledge financial support from the LFSP Nanomaterialien 
and SFB 382. We are grateful to T. Pfau for insightful discussions, and to 
G. Modugno for interesting discussions on the experiments of 
Refs.\ \cite{modugno03,ott04,pezze04}. We thank HLR-Stuttgart (Project DynMet) 
for allocation of computer time.

\end{acknowledgments}


\begin{thebibliography}{99}

\bibitem{anderson} M. H. Anderson, J. R. Ensher, M. R. Matthews, C. E.
Wieman, and E. A. Cornell, Science {\bf 269}, 198 (1995).

\bibitem{bradley} C. C. Bradley, C. A. Sackett, J. J. Tollett, and R. G.
Hulet, Phys. Rev. Lett. {\bf 75}, 1687 (1995).

\bibitem{davis} K. B. Davis, M.-O. Mewes, M. R. Andrews, N. J. van Druten,
D. S. Durfee, D. M. Kurn, and W. Ketterle, Phys. Rev. Lett. {\bf
75}, 3969 (1995).

\bibitem{ohara}
K. M. O'Hara, S. L. Hemmer, M. E. Gehm, S. R. Granade, and J. E. Thomas, 
Science {\bf 298}, 2179 (2002).

\bibitem{dalfovo} F. Dalfovo, S. Giorgini, L. P. Pitaevskii, and S.
Stringari, Rev. Mod. Phys. {\bf 71}, 463 (1999).

\bibitem{gleisberg} F. Gleisberg, W. Wonneberger, U. Schl\"oder, 
and C. Zimmermann, Phys. Rev. A {\bf 62}, 063602 (2000).

\bibitem{vignolo1} P. Vignolo, A. Minguzzi, and M. P. Tosi, 
Phys. Rev. Lett. {\bf 85}, 2850 (2000).

\bibitem{minguzzi} A. Minguzzi, P. Vignolo, and M. P. Tosi, 
Phys. Rev. A {\bf 63}, 063604 (2001) .

\bibitem{vignolo2} P. Vignolo, A. Minguzzi, and M. P. Tosi, 
Phys. Rev. A {\bf 64}, 023421 (2001).

\bibitem{schneider} J. Schneider and H. Wallis, 
Phys. Rev. A {\bf 57}, 1253 (1998).

\bibitem{brack} M. Brack and B. P. van Zyl, 
Phys. Rev. Lett. {\bf 86}, 1574 (2001).

\bibitem{vignolo3} P. Vignolo and A. Minguzzi, 
Phys. Rev. A {\bf 67}, 053601 (2003).

\bibitem{greiner} M. Greiner, O. Mandel, T. Esslinger, T. W.
H\"ansch, and I. Bloch, Nature (London) {\bf 415}, 39 (2002).

\bibitem{zoller} W. Hofstetter, J. I. Cirac, P. Zoller, 
E. Demler, and M. D. Lukin, Phys. Rev. Lett. {\bf 89}, 220407 (2002).

\bibitem{modugno03} G. Modugno, F. Ferlaino, R. Heidemann, G. Roati, 
and M. Inguscio, Phys. Rev. A {\bf 68}, 011601(R) (2003).

\bibitem{ott04} H. Ott, E. de Mirandes, F. Ferlaino, G. Roati, G. Modugno, 
and M. Inguscio, Phys. Rev. Lett. {\bf 92}, 160601 (2004).

\bibitem{pezze04} L. Pezze', L. Pitaevskii, A. Smerzi, S. Stringari, 
G. Modugno, E. de Mirandes, F. Ferlaino, H. Ott, G. Roati, 
and M. Inguscio, Phys. Rev. Lett. {\bf 93}, 120401 (2004). 

\bibitem{kennedy04} T. A. B. Kennedy, Phys. Rev. A {\bf 70}, 023603 (2004).

\bibitem{ruuska04} V. Ruuska and P. T\"orm\"a, 
New J. Phys. {\bf 6}, 59 (2004).

\bibitem{rigol03_1} M. Rigol, A. Muramatsu, G. G. Batrouni, 
and R. T. Scalettar, Phys. Rev. Lett. {\bf 91}, 130403 (2003).

\bibitem{rigol03_2} M. Rigol and A. Muramatsu, 
Phys. Rev. A {\bf 69}, 053612 (2004).

\bibitem{batrouni} G. G. Batrouni, V. Rousseau, R. T. Scalettar, M.
Rigol, A. Muramatsu, P. J. H. Denteneer and M. Troyer, Phys. Rev.
Lett. {\bf 89}, 117203 (2002).

\bibitem{zwerger} W. Zwerger, J. Opt. B: Quantum Semiclassical Opt. {\bf 5}, 
S9 (2003).

\bibitem{pfau} We thank T. Pfau for suggesting the line of thinking 
displayed below.

\bibitem{zener34} C. Zener, Proc. R. Soc. London, Ser. A {\bf 145}, 523 (1934).

\bibitem{kaskurnikov} V. A. Kashurnikov, N. V. Prokof'ev, and B. V.
Svistunov, Phys. Rev. A {\bf 66}, 031601(R) (2002).

\bibitem{mahan} G. D. Mahan, {\it Many-Particle Physics} 
(Plenum, New York and London, 1986).

\end{thebibliography}
\end{document}